\pdfoutput=1

\documentclass[11pt]{article}

\usepackage[preprint]{acl}

\usepackage{times}
\usepackage{latexsym}

\usepackage[T1]{fontenc}

\usepackage[utf8]{inputenc}

\usepackage{microtype}

\usepackage{inconsolata}

\usepackage{algorithm}
\usepackage{algorithmic}
\usepackage{xcolor}
\usepackage{pifont}
\usepackage{natbib}
\usepackage{caption}
\usepackage{graphicx}
\usepackage{booktabs}
\usepackage{multirow}
\usepackage{amsmath} 
\usepackage{amsfonts}
\usepackage{amsthm}
\usepackage{bm}
\usepackage{tcolorbox}
\usepackage{enumitem}
\usepackage{subcaption}
\usepackage{makecell}
\usepackage{subcaption}
\newcommand{\nc}[1]{{\textbf{\color{black} [\textit{\color{green}\underline{\color{purple} NEED CITE!}}]}}}

\usepackage[normalem]{ulem}
\useunder{\uline}{\ul}{}

\title{MAPS: Motivation-Aware Personalized Search via LLM-Driven Consultation Alignment}

\author{
 \textbf{Weicong Qin\textsuperscript{1}\thanks{This work was done during an internship at Lenovo
 Inc.}},
 \textbf{Yi Xu\textsuperscript{1}\footnotemark[1]},
 \textbf{Weijie Yu\textsuperscript{2}\thanks{Corresponding author}},
 \textbf{Chenglei Shen\textsuperscript{1}\footnotemark[1]},
\\
 \textbf{Ming He\textsuperscript{3}},
 \textbf{Jianping Fan\textsuperscript{3}},
 \textbf{Xiao Zhang\textsuperscript{1}},
 \textbf{Jun Xu\textsuperscript{1}},
\\
 \textsuperscript{1}Gaoling School of Artificial Intelligence, Renmin University of China, China\\
 \textsuperscript{2}University of International Business and Economics, China\\
 \textsuperscript{3}AI Lab at Lenovo Research, Lenovo Group Limited, China\\
\texttt{{yu@uibe.edu.cn}}
\\
}
\begin{document}
\maketitle
\begin{abstract}

Personalized product search aims to retrieve and rank items that match users' preferences and search intent. Despite their effectiveness, existing approaches typically assume that users' query fully captures their real motivation. However, our analysis of a real-world e-commerce platform reveals that users often engage in relevant consultations before searching, indicating they refine intents through consultations based on motivation and need. The implied motivation in consultations is a key enhancing factor for personalized search.
This unexplored area comes with new challenges including aligning contextual motivations with concise queries, bridging the category-text gap, and filtering noise within sequence history. To address these, we propose a Motivation-Aware Personalized Search (MAPS) method. It embeds queries and consultations into a unified semantic space via LLMs, utilizes a Mixture of Attention Experts (MoAE) to prioritize critical semantics, and introduces dual alignment: (1) contrastive learning aligns consultations, reviews, and product features; (2) bidirectional attention integrates motivation-aware embeddings with user preferences. Extensive experiments on real and synthetic data show MAPS outperforms existing methods in both retrieval and ranking tasks. Code is available at: \url{https://github.com/E-qin/MAPS}

\end{abstract}

\section{Introduction}

Personalized product search techniques have become crucial for e-commerce platforms and search engines~\citep{ai2019zero}, aiming to provide customized results by integrating user information during searches~\citep{bi2020transformer}.

\begin{figure}[!t]
\centering
\includegraphics[width=1\columnwidth]{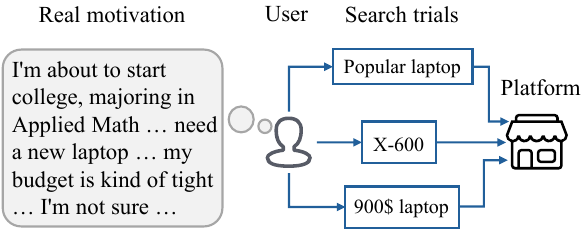}
  \caption{Illustration of the user making multiple search attempts to find the best option.}
  \label{fig:intro-motivation}
\end{figure}

Existing personalized search methods primarily focus on extracting features from users' interaction sequences to predict their interests, addressing the direct matching problem between search queries and products. These approaches include embedding-based~\citep{ai2017learning} and attention-based methods~\citep{ai2019zero}, etc. While effective, they are based on the assumption that users' queries are clear and directly describe their needs. However, in practice, we find that when initiating a search, user's queries do not always clearly articulate the requirements. For instance, as illustrated in Fig.~\ref{fig:intro-motivation}, a user searching for ``X-600'' may not be certain if this product best meets their needs and might need to conduct several searches or comparisons to find the most suitable option. 
The user's search originates from an intrinsic search motivation, which stems from the specific context or problem a user aims to resolve. Understanding and addressing such search motivations often enhances user satisfaction more effectively than merely providing products related to the query keywords. However, users typically do not actively express their motivations, which are also difficult to directly infer from query texts.

\begin{figure}[!t]
\begin{subfigure}[h]{0.5\textwidth}
        \centering
    \includegraphics[width=\textwidth]{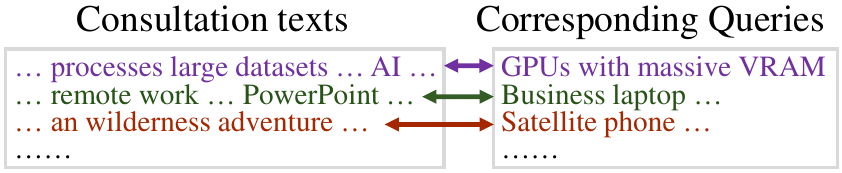}
    \caption{Example of consultations and the corresponding queries}
    \label{fig:intro:cq_example}
\end{subfigure}
\begin{subfigure}[h]{0.5\textwidth}
        \centering
    \includegraphics[width=0.85\textwidth]{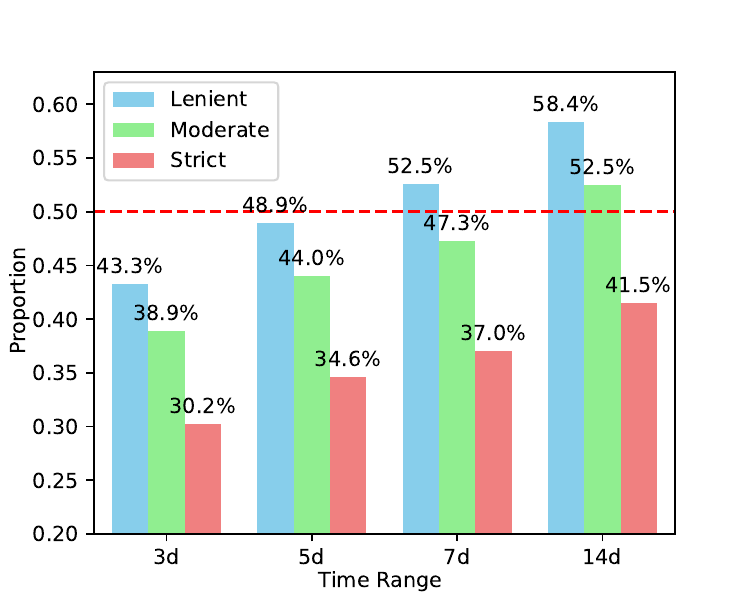}
    \caption{Proportion of search sessions with related consultations}
\end{subfigure}
  \caption{Examples of consultations with the corresponding search queries and the proportion of search sessions with related consultations (classified as ``Lenient'', ``Moderate'', or ``Strict'' under predefined NLP rules, detailed in App.~\ref{app:intro:relate:cons}) in a real e-commerce platform equipped with AI consultation services.}
  \label{fig:intro-proportion}
\end{figure}

Recently, new e-commerce platforms have developed AI consultation services that help users clarify their needs and provide guidance through natural language interactions. As shown in Fig.~\ref{fig:intro-proportion}, we conduct an experiment on a real-world e-commerce platform and find that there is a considerable probability that users conduct relevant consultations before searching, which implies the motivations behind upcoming searches. By leveraging this, the platform can appropriately adjust search results, thereby more likely fulfilling the needs driven by user search motivations, much like how personalized search utilizes users' historical search interactions to tailor search results.

For existing works, the search motivations within consultations are unexplored. It poses a new challenge: how to enable models to understand the complex texts in consultation sequences and extract motivations that align with personalized search while avoiding semantic conflicts and noise interference?

Specifically, motivation-aware search faces three critical challenges:  
(1) Alignment with Queries: Search motivations represent users' underlying needs, often involving lengthy and complex descriptions related to their personal situations, while queries typically consist of keywords directly linked to product names, models, or attributes.
(2) Alignment with Product Features: Products have categorical attributes, whereas search motivations in consultations are expressed in natural language, creating an essential disparity between the two.
(3) Alignment with User History: Not all consultation histories are relevant to the current search. The misalignment can lead to irrelevant results.

To address the challenge of aligning search motivations with diverse data sources in personalized search systems, we propose a \textbf{M}otivation-\textbf{A}ware 
\textbf{P}ersonalized \textbf{S}earch (MAPS) model. It leverages LLMs to embed queries and consulting texts into a unified semantic space and introduces a Mixture of Attention Experts (MoAE) network, which focuses on key tokens to accurately capture semantic embeddings. It also incorporates both general alignment and user-personalized alignment modules. The former extracts keyword-item relationships from diverse data sources (e.g., consulting scenarios, reviews, searches, and product features) with automated rules and contrastive learning. The latter uses bidirectional attention to extract motivation-aware embeddings from users' consulting and search histories 
and align them with individual preferences. Key contributions are as follows:
\begin{itemize}[leftmargin=*]
    \item We are the first to explicitly model ``search motivation'' and clarify its critical role within personalized search systems on e-commerce platforms that offer consulting services. 
    \item We propose MAPS, a model framework that incorporates LLM knowledge to bridge the gap between ID and text embeddings through the Mixture of Attention Experts (MoAE), effectively aligning the search motivation in personalized search modeling.
    \item Extensive experiments across retrieval and ranking stages on both a real-world commercial dataset and a synthetic dataset demonstrate that MAPS outperforms existing traditional retrieval methods, personalized search methods, and conversational retrieval methods.
\end{itemize}

\section{Related Works}

Personalized product search provides relevant items based on the user's searched queries~\citep{shi2024unisar}. 
Traditional retrieval algorithms (e.g., BM25~\cite{bm25}) primarily focus on word frequency. Dense retrieval algorithms (e.g., BGE-M3~\cite{chen2024bge}) have introduced concepts like embeddings to enhance both retrieval and ranking capabilities. Conversational retrieval methods, such as CHIQ~\cite{mo2024chiq}, have attempted to improve retrieval result accuracy by considering historical search queries. None of them provide personalized results tailored to users' interaction history or profiles. 

In recent years, personalized search has gradually attracted attention. For instance, QEM~\cite{ai2019zero} and DREM~\cite{ai2019explainable} consider the similarity between the query and items, while HEM~\cite{ai2017learning}, AEM~\cite{ai2019zero}, ZAM~\cite{ai2019zero}, and TEM~\cite{bi2020transformer} incorporate user information and search history into a separate user embedding for personalized search. Besides, several methods combine search and recommendation, such as SESRec~\cite{SESRec} using contrastive learning, UnifiedSSR~\cite{xie2023unifiedssr} with a dual-branch network for product and query histories, and UniSAR~\cite{shi2024unisar} employing transformers and cross-attention. However, these approaches inadequately model search intent and fail to align user history with semantic content




\section{Problem Formulation}

For each user $u \in U$, the chronologically ordered history $H$ consists of the search history sequence $H^\mathcal{S}$ and the consultation history sequence $H^\mathcal{C}$. Specifically, $H^\mathcal{S} = \{h^\mathcal{S}_1, \ldots, h^\mathcal{S}_N\}$, where $h^\mathcal{S}_i$ represents the $i$-th search session containing the query $s_i$ and interaction item $v_i$, denoted as $h^\mathcal{S}_i = (s_i, v_i)$. Similarly, $H^\mathcal{C}$ includes $M$ consultation sessions, given by $H^\mathcal{C} = \{c_1, \ldots, c_M\}$, where each $c_i$ contains the user inquiry and the consultant's response. Given the current query $s_{N+1}$ and the candidate item set $V_{N+1} = \{v_1, v_2, \ldots\}$ returned by the search engine, our objective is to model a ranking probability score $p(v|s_{N+1}, H, u)$ for each candidate item $v \in I_{N+1}$ based on the current query $s_{N+1}$, the user's history $H$, and the user profile $u$.

\section{Methodology}

\begin{figure*}[!t]
\centering
\includegraphics[width=0.8\textwidth]{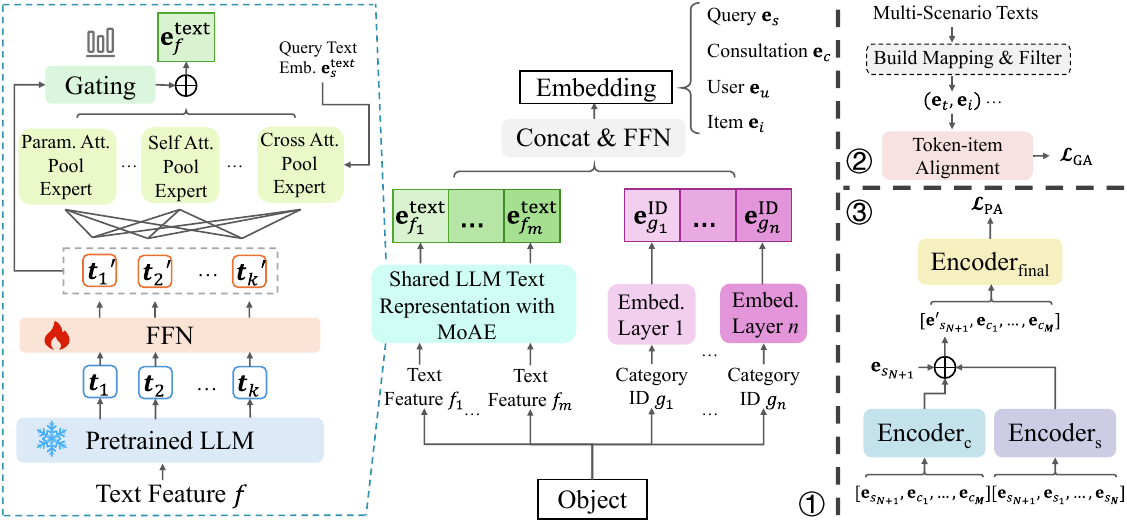}
  \caption{Overview of MAPS. \ding{172} denotes ID-text representation fusion with LLM. \ding{173} denotes the general alignment. \ding{174} denotes the personalized alignment.
  }
  \label{fig:method:big}
\end{figure*}

The overview of our MAPS is shown in Fig.~\ref{fig:method:big}.
We will introduce the following three modules in sequence:(1) ID-text representation fusion with LLM, (2) Mapping-based general alignment, and (3) Sequence-based personalized alignment.

\subsection{ID-Text Representation Fusion with LLM}

\label{sec:method-1}
In personalized product search, users, items, and various interactions need to be represented as embeddings so that the model can understand the user-item interactions. For both users and items, two types of features are included: categorical features\footnote{Such as user ID and the category; product item ID and the popularity level, etc.} and textual features.


\subsubsection{Text Representation}



Search motivations within consultations presented in natural language require models to have general natural language understanding (NLU) capabilities~\citep{allen1988natural}. Thus, it is essential to exploit the textual features of users and items fully. However, existing personalized product search methods often overlook the use of textual semantic features. For instance, regarding text, most of them initialize the token embedding layer and take average token embeddings as the query embedding~\citep{ai2017learning,ai2019explainable,ai2019zero,shi2024unisar}. This leads to models lacking world knowledge, NLU capabilities, and the ability to focus on critical semantic information within a sentence.

To address this, we consider pre-trained LLM embeddings and construct a Mixture of Attention Experts (MoAE) pooling network to adaptively assign weights to tokens in the text to acquire its embedding. Specifically, we first input the text into a frozen pre-trained LLM to get corresponding token embeddings without performing average pooling~\citep{learn}. To adapt to various LLMs, we utilize trainable feed-forward network layers (FFN) to map them to a unified dimension $d_\text{t}$. Then, we build the MoAE pooling framework, which includes three types of attention pooling experts. 







\textbf{Parameterized Attention Pooling Expert:} 
The expert maintains a parameterized embedding and utilizes it as the query within an attention mechanism. Embeddings of input tokens serve as keys to compute attention scores, resulting in a weighted average text embedding based on these scores:
\begin{equation*}
\mathbf{e}^\text{pool}_{\text{param}} = \frac{1}{L}\sum_{i=1}^{L} \text{softmax}\left(\frac{\mathbf{q}^\top (\mathbf{h}_i \mathbf{W}^k)}{\sqrt{d_\text{t}}}\right) \mathbf{h}_i,
\end{equation*}
where $\mathbf{W}^k \in \mathbb{R}^{d_\text{t} \times d_\text{t}}$ is the key matrix of the attention, $\mathbf{q}$ is the parameter query of the expert, $\mathbf{h}_i$ represents the $i$-th token embedding, and $L$ denotes the sequence length.

\textbf{ Self-Attention Pooling Expert:} 
This expert computes self-attention scores directly from the embeddings of input tokens to produce a weighted average representation:
\begin{equation*}
\mathbf{e}^\text{pool}_{\text{self}} = \frac{1}{L}\sum_{i=1}^{L} \text{softmax}\left(\frac{(\mathbf{h}_i \mathbf{W}^q)^\top (\mathbf{h}_i \mathbf{W}^k)}{\sqrt{d_\text{t}}}\right) \mathbf{h}_i,
\end{equation*}
where $\mathbf{W}^q \in \mathbb{R}^{d_\text{t} \times d_\text{t}}$ is the query matrix of the attention.

\textbf{Search-Centered Cross-Attention Pooling Expert:}
To ensure that the texts of users, items, and consultation interactions focus more on the current search query, we take the search query text\footnote{This expert will not be activated for the search query text.} embedding $\mathbf{e}^\text{text}_s$  as the attention query vector $\mathbf{q}'$:
\begin{equation*}
\mathbf{e}^\text{pool}_{\text{cross}} = \frac{1}{L}\sum_{i=1}^{L} \text{softmax}\left(\frac{(\mathbf{q}'\mathbf{W}^q)^\top (\mathbf{h}_i \mathbf{W}^k)}{\sqrt{d_\text{t}}}\right) \mathbf{h}_i.
\end{equation*}

Each type of expert has $N_E$ members, and we activate the top $K$ experts through gating scores~\citep{precise}. The gating network computes gating scores by the embeddings, then multiplying by the gating network weight matrix to transform it into a $3N_E$ dimensional vector to select the top $K$ experts. We apply softmax to obtain the gating scores of the activated top $K$ experts with $K$ pooling embedding $\mathbf{e}^\text{pool}$. The text embedding is as follows: 
\begin{equation*}
\mathbf{e}^{\text{text}} =\sum_{j=1}^{K}\ gate_j \ \mathbf{e}^\text{pool}_j,
\end{equation*}
where $gate_j$ is gating score of the $j$-th expert.

 There may be multiple textual features $f$ for 
 a user or item. The textual feature embedding for them is:
$\mathbf{e}^{\text{text}} = \operatorname{concat}(\mathbf{e}^{\text{text}}_{f_{1}},\ldots;\mathbf{e}^{\text{text}}_{f_{m}})$, where $m$ denote the number of textual features. 

\subsubsection{Categorical ID Representation}
Generally, categorical features can be converted into corresponding ID embeddings by looking up their categorical ID, i.e., $\mathbf{e}^{\text{ID}}_{g_{id}}=\operatorname{lookup}^\text{ID}_g(id)$, where $\operatorname{lookup}^\text{ID}_g$ represents the embedding lookup operation for category $g$ with the corresponding $id$.

The categorical ID embeddings for a user or an item can be obtained through concatenation as follows: 
$\mathbf{e}^{\text{ID}} = \operatorname{concat}(\mathbf{e}^{\text{ID}}_{g_{1}},\ldots;\mathbf{e}^{\text{ID}}_{g_{n}})$, where $n$ denote the number of categorical features.

\subsubsection{Overall Representations}
After obtaining $\mathbf{e}^{\text{ID}}$ and $\mathbf{e}^{\text{text}}$, we can obtain the overall embedding for user $u$, item $v$, query $s$, and consultation $c$ by further concatenation, feed-forwarding and activation as follows:
\begin{equation}
\begin{aligned}
   & \mathbf{e}_u = \operatorname{act}(\operatorname{FFN}_\text{u}(\operatorname{concat}(\mathbf{e}^{\text{ID}}_u,\mathbf{e}^{\text{text}}_u)), \\
& \mathbf{e}_v = \operatorname{act}(\operatorname{FFN}_\text{v}(\operatorname{concat}(\mathbf{e}^{\text{ID}}_v,\mathbf{e}^{\text{text}}_v)),\\
&   \mathbf{e}_{s} =\operatorname{act}(\operatorname{FFN}_\text{s}(\mathbf{e}^\text{text}_{s})), \\
&\mathbf{e}_{c} = \operatorname{act}(\operatorname{FFN}_\text{c}(\mathbf{e}^\text{text}_{c})).
\end{aligned} 
\label{eq:act}
\end{equation}
where $\operatorname{FFN}_\text{k}()$ with $k \in \{\text{u},\text{v},\text{s},\text{c}\}$ is used to map embeddings to a unified dimension $d_{\text{uni}}$, and $\operatorname{act}$ refers to the activation function.

\subsection{Mapping-Based General Alignment}
\label{sec:method-2}


To enable the model to understand which features IDs and items correspond to various texts, it is crucial to align tokens and items within a unified semantic space, referred to as ``general alignment'' in this paper. For an item $v$, we gather all relevant textual data across multiple scenarios involving this item, including related query sets, consultations, item titles, descriptive texts, advertisement texts, etc., thereby constructing a comprehensive full-text collection $\mathcal{A}_v$ for item $v$. 

We further refine the keyword collection by setting a threshold $t$ to filter out noise texts that occur infrequently in search-related text scenarios:
\begin{equation}
\mathcal{A}^{\mathcal{S}}_v = \operatorname{filter}^\mathcal{S}(\mathcal{A}_v) = \{w \in \mathcal{A}_v | \text{freq}^\mathcal{S}(w) > t\},
\label{eq:threshold}
\end{equation}
where $\text{freq}^\mathcal{S}(w)$ represents the frequency of word $w$ in search-related scenarios $\mathcal{S}$. This curated collection allows us to establish a mapping $M$ from tokens to items, where each token $t \in \mathcal{A}^{\mathcal{S}}_v$ is associated with item $v$ through their shared presence in similar search contexts or thematic relevance. Given the mapping $M$, we calculate a similarity score between token $t$ and item $v$ as $\operatorname{sim}(\mathbf{e}_t, \mathbf{e}_v)$, where $\operatorname{sim}(\cdot)$ represents the dot product similarity function.

For embeddings in a shared semantic space, given token-item pairs $(t,v)$, we introduce a bidirectional contrastive loss $\mathcal{L}_{\text{GA}}$ as follows:
\begin{equation*}
\begin{aligned}
    \mathcal{L}_{\text{GA}} = 
  &  - \lambda_1 \sum_{(t,v)}\log \frac{\exp(\operatorname{sim}(\mathbf{e}_t, \mathbf{e}_v) / \tau_1)}{\sum_{t^- \in T_{\text{neg}}} \exp(\operatorname{sim}(\mathbf{e}_{t^-}, \mathbf{e}_v) / \tau_1)} \\
   & - \lambda_2 \sum_{(t,v)}\log \frac{\exp(\operatorname{sim}(\mathbf{e}_t, \mathbf{e}_v) / \tau_2)}{\sum_{v^- \in I_{\text{neg}}} \exp(\operatorname{sim}(\mathbf{e}_t, \mathbf{e}_{v^-}) / \tau_2)},
\end{aligned}
\end{equation*}
where $I_{\text{neg}}$ and $T_{\text{neg}}$  are respectively the set of randomly sampled negative items and tokens,  $\lambda_1$ and $\lambda_2$ are weights for the two terms, and $\tau_1$, $\tau_2$ are temperature parameters for controlling the sharpness of the softmax distribution~\citep{hinton2015distilling}. This formulation ensures that the model learns to assign higher similarity scores to correct token-item pairs compared to incorrect ones.

\subsection{Sequence-Based Personalized Alignment}

This section shows how to mine the search motivations in consultations and align it to the current query $s_{N+1}$ to enhance searching. 

\subsubsection{\mbox{Motivation-Aware Query Embedding}}
Inspired by~\citet{bi2020transformer}, we treat the embedding of the current query $\mathbf{e}_{s_{N+1}}$ as an anchor. It, together with the user's consultation history $[\mathbf{e}_{c_1}; \ldots; \mathbf{e}_{c_M}]$, is fed into an  transformer encoder. Through a multi-head bidirectional attention mechanism and FFN layers, we obtain the search motivation embedding from the user's consultation history. The formula is as follows:
\begin{equation*}
    \mathbf{e}^\mathcal{C}_{s_{N+1}} = \text{Encoder}_{\text{c}}( \mathbf{e}_{s_{N+1}},\mathbf{e}_{c_1}, \ldots, \mathbf{e}_{c_M})[0,:],
\end{equation*}
where $[0,:]$  denotes the operation of selecting the first vector. $\mathbf{e}^\mathcal{C}_{s_{N+1}}$ denotes the search motivation from consultation history for the current query.

Considering that historical query may also have certain relevance or implications for the motivation of the current query~\citep{SESRec}, we perform the same operation on the query history $[\mathbf{e}_{s_1}; \ldots; \mathbf{e}_{s_M}]$. The formula is as follows:
\begin{equation*}
    \mathbf{e}^\mathcal{S}_{s_{N+1}} = \text{Encoder}_{\text{s}}( \mathbf{e}_{s_{N+1}},\mathbf{e}_{s_1}, \ldots, \mathbf{e}_{s_M})[0,:],
\end{equation*}
where $\mathbf{e}^\mathcal{S}_{s_{N+1}}$ denotes the search motivation from query history for the current query.

Furthermore, we obtain the aggregated motivation-aware query embedding $\mathbf{e}'_{s_{N+1}}$ through learnable weights as follows:
\begin{equation*}
    \mathbf{e}'_{s_{N+1}} = \alpha_1 \mathbf{e}^\mathcal{C}_{s_{N+1}} + \alpha_2\mathbf{e}^\mathcal{S}_{s_{N+1}} + \alpha_3\mathbf{e}_{s_{N+1}},
\end{equation*}
where $\alpha_1, \alpha_2$ and $\alpha_3$  are learnable weights. 

\subsubsection{Personalized Search with Item History}

Given the motivation-aware query embedding $\mathbf{e}'_{s_{N+1}}$ and item embeddings $\mathbf{E}_\text{items}$, we input them into a transformer encoder to capture complex interactions, adding the user embedding $\mathbf{e}_u$ to obtain the final query embedding $\mathbf{e}''_{s_{N+1}}$:
\begin{equation*}
    \mathbf{e}''_{s_{N+1}} = \text{Encoder}_{\text{final}}(\mathbf{e}'_{s_{N+1}}, \mathbf{E}_\text{items})[0,:] \oplus \mathbf{e}_u,
\end{equation*}
where $\oplus$ denotes the in-place add for the vector.

For inference, the candidate items can be ranked based on their similarity-derived probability scores:
\begin{equation*}
p(v|s_{N+1},H, u) =  \operatorname{sim}( \mathbf{e}''_{s_{N+1}},\mathbf{e}_v),
\end{equation*}
where $v \in V_{N+1}$ is the candidate item, $\operatorname{sim}(\cdot)$ denotes by the dot product function.  

In terms of optimization, following previous methods~\citep{bi2020transformer,ai2017learning,shi2024unisar}, our learning objective is to increase the relevance scores of ground-truth given user sequence. The personalized alignment loss $\mathcal{L}_\text{PA}$ can be formulated as:
\begin{equation*}
    \mathcal{L}_\text{PA} = \sum_{(u,v,s_{N+1})} \log \frac{\exp(\operatorname{sim}(\mathbf{e}''_{s_{N+1}},\mathbf{e}_{v}))}{\sum_{v' \in V_{N+1}} \exp(\operatorname{sim}( \mathbf{e}''_{s_{N+1}},\mathbf{e}_{v'}))}.
\end{equation*}
Following existing works~\cite{ai2019explainable,shi2024unisar}, we employ negative sampling~\citep{le2014distributed} for $\mathcal{L}_\text{PA}$. The overall loss $\mathcal{L}_{\text{overall}}$ is:
\begin{equation*}
\mathcal{L}_{\text{overall}} = \mathcal{L}_\text{PA} + \lambda_3 \mathcal{L}_\text{GA} + \lambda_4 ||\Theta||_2,
\end{equation*}
where $\lambda_3$ and $\lambda_4$ is a hyper-parameters, $\Theta$ is the parameters of MAPS.


\section{Experiment}
 We answer the following research questions in this section: 
\textbf{RQ1:} How does MAPS perform compared to existing baselines regarding ranking performance? 
\textbf{RQ2:} How effective is MAPS regarding retrieval?  
\textbf{RQ3:} How does MAPS perform compared to multi-scenario methods? 
\textbf{RQ4:} How effective is each module introduced by MAPS?
\textbf{RQ5:} Does MAPS exhibit scalability?  
\textbf{RQ6:} How do integrated ID and LLM embeddings with MoAE pooling enhance personalized search?

\subsection{Experiment Settings}
\subsubsection{Datasets} 

To validate the effectiveness of MAPS, we conduct experiments on two datasets. 
\textbf{Commercial dataset} is a real user interaction dataset from an internet e-commerce shopping platform with AI consulting services with interactions in 31 days. We filter out users and items with fewer than 5 interactions, following~\citep{FMLPREC,shi2024unisar}. To prevent sequence data leakage~\citep{ji2023critical}, we select the first 29 days for training, with the remaining two days used respectively for validation and testing. \textbf{Amazon}: To validate users' search motivations and leverage LLM knowledge, a dataset with real token text\footnote{Datasets like KuaiSAR~\citep{kuaisar} and Jdsearch~\citep{jdsearch} only provide encrypted token texts, making it impossible to verify the capability of understanding motivations based on natural language. Therefore, they are not suitable for our experiments.} and multiple types of user interaction data is essential. Therefore, we adopt the widely used Amazon Reviews dataset~\citep{ni2019justifying}. Specifically, we use the version processed by PersonalWAB~\citep{cai2024personalwab}, which includes user profiles and various types of user interactions such as searches and reviews. To simulate the real-world e-commerce platform with AI consultation services, we utilized GPT-4o\footnote{\url{https://openai.com}} to generate user consultation texts based on user profiles and interaction behaviors. The processing and splitting are kept consistent with~\citet{shi2024unisar}. The statistics of these datasets are shown in Tab.~\ref{tab:exp:statistics}.

\begin{table}[!tbp]
\centering\setlength{\tabcolsep}{2.2pt}\fontsize{8}{7}\selectfont
\begin{tabular}{lccccc}
\toprule
\textbf{Dataset}         & \textbf{\#Users}   & \textbf{\#Items}   & \textbf{\#Inters}   & \textbf{\#Sparsity}   \\ \midrule
\textbf{Commercial}    & 2096           & 2691  & 24662, (18774)           & 99.56\%, (99.66\%)          \\
\textbf{Amazon}   & 967           & 35772 & 7263, (40567)            & 99.98\%, (99.88\%)          \\
\bottomrule
\end{tabular}
\caption{Statistics of the 2 pre-processed datasets. In ``\#Inters'' and ``\#Sparsity'', the numbers in parentheses indicate consultation interactions, while the numbers outside the parentheses indicate search interactions.}
\label{tab:exp:statistics}
\end{table}

\subsubsection{Baselines}
For \textbf{ranking} performance comparison, we adopt these personalized search baselines: \textbf{ZAM}~\citep{ai2019zero}, \textbf{HEM}~\citep{ai2017learning}, \textbf{AEM}~\citep{ai2019zero}, \textbf{QEM}~\citep{ai2019zero}, \textbf{TEM}~\citep{bi2020transformer}, and \textbf{CoPPS}~\citep{CoPPS}. For \textbf{retrieval} performance, MAPS is compared with traditional, dense, and conversational retrieval methods: \textbf{BM25}~\citep{bm25}, \textbf{BGE-M3}~\citep{chen2024bge}, and \textbf{CHIQ}~\citep{mo2024chiq}. Additionally, we also consider multi-scenario baselines that integrate search and recommendation interactions for better ranking, including \textbf{SESRec}~\citep{SESRec}, \textbf{UnifiedSSR}~\citep{xie2023unifiedssr}, and \textbf{UniSAR}~\citep{shi2024unisar}. More details can be found in App.~\ref{app:exp:baselines}.

\subsubsection{Metrics and Implementation details}
Consistent with previous works~\cite{SESRec,shi2024unisar}, we adopt \textbf{Hit Ratio} (HR@$k$) and \textbf{Normalized Discounted Cumulative Gain} (NDCG@$k$ or N@$k$) for ranking metrics. For retrieval evaluation, we adopt \textbf{Mean Reciprocal Rank} (MRR@$k$). Following ~\citet{shi2024unisar}, we pair the ground-truth item with 99 randomly sampled negative items as candidates and report HR and NDCG at $\{5,10, 20, 50\}$ for ranking evaluation. For retrieval, we consider all items as candidates and report MRR at $\{10, 20, 50\}$ for the retrieval task. For more model settings and implementation details, see App.~\ref{app:exp:implement}.


\begin{table*}[t]
\centering
\small
\begin{tabular}{ccccccccc}
\toprule
\multicolumn{1}{c|}{Model} & HR@5            & HR@10           & HR@20           & HR@50           & NDCG@5          & NDCG@10         & NDCG@20         & NDCG@50         \\ \midrule
\multicolumn{9}{c}{Commercial}                                                                                                                                             \\ \midrule
\multicolumn{1}{c|}{AEM}   & 0.3886          & 0.5376          & 0.6733          & 0.8249          & 0.2656          & 0.3135          & 0.3478          & 0.3781          \\
\multicolumn{1}{c|}{QEM}   & 0.3996          & 0.5473          & 0.6733          & 0.8439          & 0.2671          & 0.3144          & 0.3463          & 0.3805          \\
\multicolumn{1}{c|}{HEM}   & 0.3484          & 0.4907          & 0.6366          & 0.8037          & 0.2360          & 0.2817          & 0.3185          & 0.3519          \\
\multicolumn{1}{c|}{ZAM}   & 0.3674          & 0.5248          & 0.6808          & 0.8205          & 0.2490          & 0.2994          & 0.3389          & 0.3669          \\
\multicolumn{1}{c|}{TEM}   & 0.4041          & 0.5685          & 0.7078          & 0.8528          & 0.2871          & 0.3402          & 0.3756          & 0.4049          \\
\multicolumn{1}{c|}{CoPPS} & 0.4050          & 0.5637          & 0.7171          & 0.8660          & 0.2831          & 0.3445          & 0.3805          & 0.4103          \\
\multicolumn{1}{c|}{MAPS}  & \textbf{0.5281}$^\dagger$ & \textbf{0.7071}$^\dagger$ & \textbf{0.8330}$^\dagger$ & \textbf{0.9308}$^\dagger$ & \textbf{0.3780}$^\dagger$ & \textbf{0.4359}$^\dagger$ & \textbf{0.4680}$^\dagger$ & \textbf{0.4877}$^\dagger$ \\ \midrule
\multicolumn{9}{c}{Amazon}                                                                                                                                            \\ \midrule
\multicolumn{1}{c|}{AEM}   & 0.3180          & 0.4550          & 0.5372          & 0.7239          & 0.1860          & 0.2132          & 0.2475          & 0.2768          \\
\multicolumn{1}{c|}{QEM}   & 0.2831          & 0.3888          & 0.5285          & 0.7663          & 0.1914          & 0.1805          & 0.2277          & 0.2913          \\
\multicolumn{1}{c|}{HEM}   & 0.2735          & 0.4198          & 0.5400          & 0.7446          & 0.1983          & 0.2172          & 0.2598          & 0.2961          \\
\multicolumn{1}{c|}{ZAM}   & 0.3103          & 0.4488          & 0.5429          & 0.7301          & 0.1833          & 0.2114          & 0.2494          & 0.2787          \\
\multicolumn{1}{c|}{TEM}   & 0.4026          & 0.4814          & 0.7197          & 0.7301          & 0.2968          & 0.3124          & 0.3415          & 0.3535          \\
\multicolumn{1}{c|}{CoPPS} & 0.3870          & 0.4854          & 0.7286          & 0.8004          & 0.2788          & 0.3298          & 0.3439          & 0.3699          \\
\multicolumn{1}{c|}{MAPS}  & \textbf{0.5832}$^\dagger$ & \textbf{0.7735}$^\dagger$ & \textbf{0.8987}$^\dagger$ & \textbf{0.9741}$^\dagger$ & \textbf{0.4059}$^\dagger$ & \textbf{0.4676}$^\dagger$ & \textbf{0.4995}$^\dagger$ & \textbf{0.5147}$^\dagger$ \\ \bottomrule
\end{tabular}
\caption{Search ranking performance compared with personalized search baselines. The best results are shown in bold. `$\dagger$' indicates the model significantly outperforms all baseline models with paired t-tests at $p < 0.05$ level.}
\label{tab:exp:ranking}
\end{table*}

\begin{table}[t]
\centering
\small
\begin{tabular}{c|ccc}
\toprule
Method           & MRR@10          & MRR@20         & MRR@50          \\ \midrule
BM25             & 0.2529          & 0.2577         & 0.2625          \\
AEM              & 0.2445          & 0.2539         & 0.2588          \\
QEM              & 0.2427          & 0.2516         & 0.2572          \\
HEM              & 0.2176          & 0.2277         & 0.2331          \\
ZAM              & 0.2304          & 0.2413         & 0.2459          \\
TEM              & 0.2705          & 0.2803         & 0.2852          \\
CoPPS            & 0.2642          & 0.2750         & 0.2799          \\
BGE-M3           & 0.2976         & 0.3110        & 0.3168          \\
CHIQ             & 0.3192          & 0.3392         & 0.3412          \\ \midrule
MAPS             & \textbf{0.3805  } & \textbf{0.3889 } & \textbf{0.3922} \\ \bottomrule
\end{tabular}
\caption{Retrieval performance on the Commercial dataset.}
\label{tab:exp:retrieval}
\end{table}

\begin{table}[t]
\centering
\small
\begin{tabular}{l|cccc}
\toprule
Method   & HR@10           & HR@20           & N@10         & N@20         \\ \midrule
SESRec       & 0.5622          & 0.7191          & 0.3465          & 0.3797          \\
UnifiedSSR     & 0.5706          & 0.7074          & 0.3590          & 0.3743          \\
UniSAR        & 0.5838          & 0.7294          & 0.3577          & 0.3894         \\
\midrule
MAPS               & \textbf{0.7071} & \textbf{0.8330} & \textbf{0.4359} & \textbf{0.4680} \\
\bottomrule
\end{tabular}
\caption{Search ranking performance compared with multi-scenario baselines on the Commercial dataset.}
\label{tab:exp:cross:comp}
\end{table}

\begin{table}[t]
\centering
\setlength{\tabcolsep}{2.5pt}
\small
\begin{tabular}{l|cccc}
\toprule
Ablation   & HR@10           & HR@20           & N@10         & N@20         \\ \midrule
MAPS               & \textbf{0.7071} & \textbf{0.8330} & \textbf{0.4359} & \textbf{0.4680} \\ \midrule
\ w/o LLM            & 0.6527          & 0.7839          & 0.3968          & 0.4309          \\
\ w/o MoAE           & 0.6781          & 0.7844          & 0.4096          & 0.4494          \\ \midrule
\ w/o general align  & 0.6198          & 0.7424          & 0.3669          & 0.4006          \\
\quad w/o $\operatorname{filter}^\mathcal{S}()$ in Eq.~\ref{eq:threshold}  & 0.6201  & 0.7426 & 0.3597 & 0.3951 \\ \midrule
\ w/o personal align & 0.6334          & 0.7518           & 0.3732   & 0.4105     \\
\quad  w/o $\mathbf{e}^\mathcal{S}_{s_{N+1}}$ & 0.6565          & 0.7730           & 0.3863          & 0.4246 
\\ \quad  w/o $\mathbf{e}^\mathcal{C}_{s_{N+1}}$ & 0.6448          & 0.7615           & 0.3803          & 0.4170
\\ \bottomrule
\end{tabular}
\caption{Ablation study of MAPS on the Commercial dataset.}
\label{tab:exp:ablation}
\end{table}





\subsection{Overall Performance}  
Most e-commerce platforms utilize a retrieval-then-ranking pipeline. For personalized product search, the primary objective is personalized ranking. To answer \textbf{RQ1}, \textbf{RQ2}, and \textbf{RQ3}, we first evaluate the ranking performance of various baselines and MAPS in Tab.~\ref{tab:exp:ranking} and Tab.~\ref{tab:exp:cross:comp}, followed by a comparison of the retrieval performance in Tab.~\ref{tab:exp:retrieval} with retrieval baselines. 

Regarding ranking, MAPS significantly outperforms all other personalized product search methods and search-integrated multi-scenario approaches, achieving approximately 20\% improvement on Commercial and 35\% on Amazon.
Concerning retrieval, MAPS surpasses all personalized product search methods and traditional, dense, and conversational retrieval approaches, achieving an improvement of over 15\%.
This fully demonstrates the effectiveness and superiority of the MAPS approach in both ranking and retrieval tasks, highlighting its capability to enhance search performance on e-commerce platforms.


\subsection{Ablation Study}  
In this section, we discuss the specific roles of each module in MAPS to answer \textbf{RQ4}. ``w/o personal align'' indicates to set $\mathbf{e}'_{q_{N+1}}=\mathbf{e}_{q_{N+1}}$, deleting the motivation capturing and alignment from consultation and query sequence history. From Tab.~\ref{tab:exp:ablation}, we observe that removing any module significantly reduces MAPS's performance, such as not integrating large model semantic information or not adding MoAE. The table shows that the performance drop is most pronounced when the general alignment module is not included. The general alignment specifically aligns text with IDs, as well as with scenario-specific knowledge and world knowledge in the Commercial dataset. Due to the clear gap between practical vertical retrieval-ranking scenarios and the world knowledge in LLMs, not using general alignment causes significant misalignment in embeddings and term usage. For example, if ``Cool'' in the scenario is not aligned, LLM will interpret it as an adjective rather than a product feature, resulting in a significant semantic shift and a noticeable drop in ranking performance.

\subsection{Scalability Study}  
To answer \textbf{RQ5}, we analyze MAPS's scalability from various aspects, including the used interaction sequence length during training, the LLM model and its corresponding parameter count, and the number of transformer layers used. Based on the results in Tab.~\ref{tab:exp:scalability}, we find that:  
(1) The sequence length used during training is not always better when longer; this indicates that longer user sequences may contain more noise, which can affect the final ranking performance.  
(2) The stronger the LLM's performance and the more parameters it has, the better MAPS performs. This suggests that a larger amount of world knowledge can enhance the semantic information covered by LLM embeddings, thus improving the model's ranking ability.  
(3) The more transformer layers used, the stronger the ranking effect, suggesting that multiple transformer layers can effectively align LLM embeddings with the scenario.

\subsection{ID-text Representation Fusion Analysis}  
To answer \textbf{RQ6}, we conduct an experiment on the Commercial dataset, processing the representations of users and items in different forms. The results are shown in  Tab.~\ref{tab:exp:feature}. We find that:  
(1) Fusing original category ID and LLM embeddings better represents information of user and item. 
(2) MoAE, which covers multiple attention mechanisms, can adaptively select the best attention expert to calculate attention scores for the corresponding text, allowing the final semantic embedding to better align with the search task.

\begin{table}[!t]
\centering
\setlength{\tabcolsep}{3.5pt}
\small
\begin{tabular}{l|cccc}
\toprule
     Ablation     & HR@10           & HR@20           & N@10         & N@20         \\ \midrule
MAPS-Default     & \textbf{0.7071} & \textbf{0.8330} & \textbf{0.4359} & \textbf{0.4680} \\
MAPS-ID   & 0.6870          & 0.7953          & 0.4226          & 0.4500          \\
MAPS-LLM  & 0.6794          & 0.7896          & 0.4196          & 0.4427          \\
MAPS-Mean & 0.6950          & 0.8249          & 0.4337          & 0.4566          \\ \bottomrule
\end{tabular}
\caption{The performance of representation for users and items under different settings on Commercial. `ID' denotes using only ID embedding (including categorical features), `LLM' indicates using only LLM embedding (containing text features only), and `Mean' refers to conducting mean pooling only.}
\label{tab:exp:feature}
\end{table}

\begin{table}[!t]
\centering
\setlength{\tabcolsep}{2.8pt}
\small
\begin{tabular}{c|c|ccc}
\toprule
Aspect                             & Config           & N@5 & N@10 & N@20 \\ \midrule
\multirow{3}{*}{\makecell{Sequence\\Length}}   & 10               & 0.3674 & 0.4200  & 0.4481  \\
                                   & {\ul 30}         & \textbf{0.3780} & \textbf{0.4359}  & \textbf{0.4680}  \\
                                   & 40               & 0.3739 & 0.4303  & 0.4627  \\ \midrule
\multirow{4}{*}{\makecell{LLM\\Scale}}         & Qwen2.5-0.5B     & 0.3394 & 0.3892  & 0.4237  \\
                                   & Qwen2.5-1.5B     & 0.3534 & 0.4026  & 0.4357  \\
                                   & Qwen2-7B         & 0.3593 & 0.4090  & 0.4412  \\
                                   & {\ul Qwen2.5-7B} & \textbf{0.3780} & \textbf{0.4359}  & \textbf{0.4680}  \\ \midrule
\multirow{3}{*}{\makecell{Transformer\\Scale}} & {\ul 1 Layer}    & 0.3780 & 0.4359  & 0.4680  \\
                                   & 2 Layer          & 0.3881 & 0.4470  & 0.4724  \\
                                   & 4 Layer          & \textbf{0.3909} & \textbf{0.4561}  & \textbf{0.4838}  \\ \bottomrule
\end{tabular}
\caption{Scalability Study of MAPS on the Commercial dataset. Default configurations are underlined.}
\label{tab:exp:scalability}
\end{table}

\begin{figure}[!t]
  \includegraphics[width=0.49\linewidth]{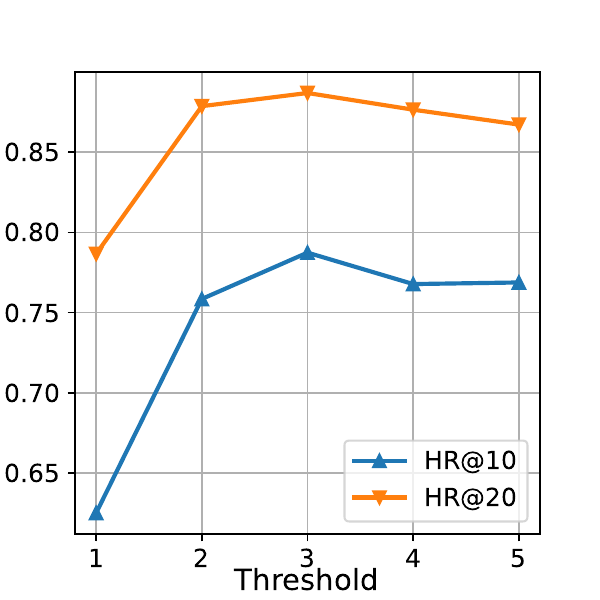} \hfill
  \includegraphics[width=0.49\linewidth]{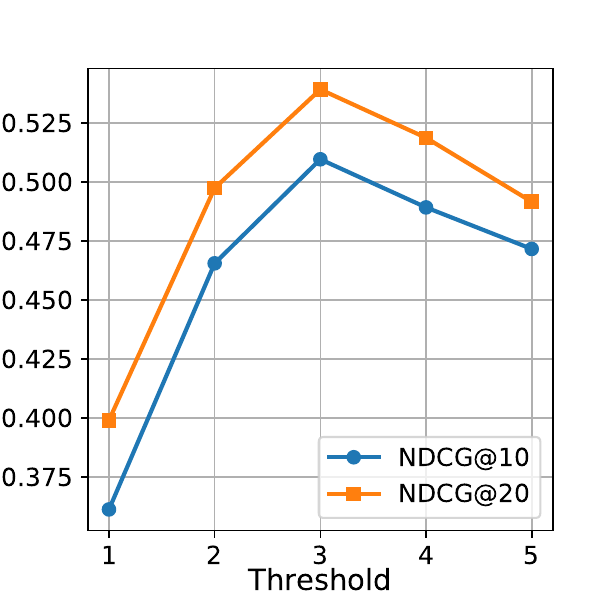}
  \caption {Ranking performance on Amazon with different threshold $t$ in Eq.~\ref{eq:threshold}. The default one is 2.}
  \label{fig:exp:threshold}
\end{figure}

\begin{table}[!t]
\centering
\small\setlength{\tabcolsep}{3.5pt}
\begin{tabular}{l|cccc}
\toprule
Activation  & \multicolumn{1}{l}{HR@10} & \multicolumn{1}{l}{HR@20} & \multicolumn{1}{l}{N@10} & \multicolumn{1}{l}{N@20} \\ \midrule
{tanh}  & 0.7585                    & 0.8787                    & 0.4676                                              & 0.4995                                              \\
SiLU  & 0.7823                    & 0.8953                    & 0.4697                                              & 0.5010                                              \\
PReLU & 0.7813                    & \textbf{0.9067}                    & \textbf{0.4763}                                              & \textbf{0.5097}                                              \\
GELU  & \textbf{0.7978}                    & 0.9036                    & 0.4734                                              & 0.5015                                              \\
ReLU  & 0.4390                    & 0.6740                    & 0.2165                                              & 0.2768                                              \\ \bottomrule
\end{tabular}
\caption {Performance on Amazon with different activation function in Eq.~\ref{eq:act}. The default one is ``tanh''.}
  \label{tab:exp:act}
\end{table}

\subsection{Configuration Analysis} 

In this section, we investigate the impact of changing the mapping threshold $t$ in Eq.~\ref{eq:threshold} and the activation function in Eq.~\ref{eq:act}. 

Fig.~\ref{fig:exp:threshold} illustrates that both excessively small and overly large values of $t$ result in performance degradation and setting $t=3$ achieves optimal performance. It suggests that a too-low threshold introduces noise from texts in other scenarios, whereas an overly high threshold imposes stringent conditions that limit the amount of useful data, thereby constraining performance. Thus, an appropriate threshold $t$ is necessary.

Following~\citet{shi2024unisar}, we use tanh as the default activation function. However, Tab.~\ref{tab:exp:act} indicates that other options seem to work better, except for ReLU. We attribute this to the ``dying ReLU''~\citep{lu2019dying}, where negative inputs cause zero gradients, thereby preventing weight updates.





\section{Conclusion}
We propose the Motivation-Aware Personalized Search (MAPS) method to enhance personalized product search by incorporating user consultations, mining search motivations to understand user intent, and aligning these insights with search systems. MAPS uses LLMs and a Mixture of Attention Experts (MoAE) to align queries and consultations in a unified semantic space. The dual alignment approaches (contrastive learning and bidirectional attention) bridge the ID-text gap, prioritize key semantics, and integrate user preferences. Experimental results show that MAPS outperforms existing methods in retrieval and ranking tasks, offering a more accurate, context-aware solution. In the future, we will explore further consultation modeling within e-commerce platforms.

\section{Limitations}
While our method significantly improves personalized search by enhancing semantic understanding through consultations, there are several areas that could be further optimized. First, while the approach effectively aligns queries and consultations, it focuses primarily on semantic alignment and may not fully address other potential bottlenecks in the search process, such as computational efficiency or scalability in real-time applications. Additionally, the current framework does not explicitly account for dynamic user behavior over time, which could affect long-term user preferences and search intent refinement. Furthermore, MAPS does not incorporate explicit domain-specific knowledge that could enhance understanding in specialized contexts, limiting its generalizability across diverse industries. Future work could focus on optimizing these aspects, such as improving real-time adaptability, addressing scalability issues, and integrating external domain knowledge for more robust and versatile personalized search systems.

\bibliography{custom}
\newpage
\appendix


\section{Appendix: Details of Related Consultations}
\label{app:intro:relate:cons}

To preliminarily validate the experiments introduced, we predefined certain conditions and rules to find relevant consultation interactions within a time period $time$ before a search session for its search, where $time \in \{3d, 5d, 7d, 14d\}$. For each query and consultation pair $(c,s)$ prior to the search timestamp, $c$ is considered a relevant consultation for $s$ if it meets the following criteria. The conditions are categorized into three levels:

To preliminarily validate the experiments introduced, we predefined certain conditions and rules to find relevant consultation interactions within a time period $time$ before a search session for its search, where $time \in \{3d, 5d, 7d, 14d\}$. For each query and consultation pair $(c,s)$ prior to the search timestamp, $c$ is considered a relevant consultation for $s$ if it meets the following criteria. The conditions are categorized into three levels:

\begin{enumerate}
    \item \textbf{Lenient}: If any interaction item $v$'s title in the search session $s$ appears in full at least once in $c$, or any category attribute of $v$ appears as text in $c$ at least once, or terms from the query of $s$ appear in $c$ at least twice.
    
    \item \textbf{Moderate}: If any interaction item $v$'s title in the search session $s$ appears in full at least once in $c$, or more than half of the attributes of $v$ appear as text in $c$, or more than half of the terms from the query of $s$ appear in $c$.
    
    \item \textbf{Strict}: If any interaction item $v$'s title in the search session $s$ appears in full at least once in $c$, or three-quarters or more of the attributes of $v$ appear as text in $c$, or three-quarters or more of the terms from the query of $s$ appear in $c$.
\end{enumerate}

These criteria allow us to evaluate the relevance of consultations to searches under different stringency levels.

\section{Appendix:  Experiment Details}


\subsection{Baseline Details}
\label{app:exp:baselines}
We begin by comparing our method with the following \textbf{personalized search models} for both ranking and retrieval tasks:
\begin{itemize}
    \item \textbf{AEM}~\cite{ai2019zero}, an attention-based personalized model that combines the user's previously interacted items with the current query.
    \item \textbf{QEM}~\cite{ai2019zero}, which only takes into account the matching scores between items and queries.
    \item \textbf{HEM}~\cite{ai2017learning}, a personalized model based on latent vectors.
    \item \textbf{ZAM}~\cite{ai2019zero}, which enhances AEM by concatenating a zero vector to the item list.
    \item \textbf{TEM}~\cite{bi2020transformer}, which improves AEM by replacing its attention layer with a transformer encoder.
    \item \textbf{CoPPS}~\cite{CoPPS}, which leverages contrastive learning techniques.
\end{itemize}

We then compare our method with multi-scenario methods that integrate search and recommendation interactions, including:
\begin{itemize}
    \item \textbf{SESRec}~\cite{SESRec} utilizes contrastive learning to learn disentangled search representations for recommendation.
    \item \textbf{UnifiedSSR}~\cite{xie2023unifiedssr} jointly learns user behavior history across both search and recommendation scenarios.
    \item \textbf{UniSAR}~\cite{shi2024unisar} effectively models the different types of fine-grained behavior transitions through two different transformers and implements a cross-attention mechanism.
\end{itemize}

For retrieval, we compare MAPS with traditional, deep learning-based, and conversational-based baselines, including:
\begin{itemize}
    \item \textbf{BM25} uses word frequency to maintain the related retrieval candidates.
    \item \textbf{BGE-M3} introduces concepts like embedding to enhance the performance of retrieval tasks.
    \item \textbf{CHIQ} attempts to incorporate world knowledge from LLM into the search for the enhancement of retrieval.
\end{itemize}

\subsection{Implementation Details}
\label{app:exp:implement}

All hyperparameters of the baseline are searched according to the settings in the original paper. 
Following related work~\citep{shi2024unisar}, we set $d$ to 64, $d_\text{t}$ to 32, and the maximum length of the user history sequence to 30. We filter out users with fewer than 5 interactions, used tanh as the activation function, set the number of layers in the transformer encoder to 1, batch size to 72, and the number of negative samples for each positive sample for $\mathcal{L}_\text{PA}$ to 10. For $\mathcal{L}_\text{GA}$, the in-batch negative is adopted, with the corresponding batch size searched among \{128,256,512,1024\}. $(\lambda_1,\lambda_2)$ is tuned in \{(0.0,1.0), (0.25,0.75), (0.5,0.5), (0.75,0.25), (1.0,0.0)\}, and $\lambda_3$ is tuned in \{0.0,0.05,0.1,0.2,0.3,0.4,0.5\}. The temperatures $\tau_1$ and $\tau_2$ are tuned in the interval [0.0, 1.0] with a step size of 0.1.
We train all models for 100 epochs using early stopping to avoid overfitting and optimize using Adam~\cite{kingma2014adam}.
The learning rate is adjusted among \{1e-3, 5e-4, 1e-4, 5e-5, 1e-5\}.
All experiments were completed on an A800 (80GB) GPU. 
\section{Appendix: Dataset License}

The Amazon dataset (based on PersonalWAB~\citep{cai2024personalwab}) is licensed under the CC BY-NC 4.0 License, its benchmark implementation, based on the MIT-licensed tau-bench~\citep{yao2024tau} with substantial modifications and enhancements for the project's needs, still respects and follows the licensing terms of tau-bench for the derived parts.

\section{Appendix: Dataset Details}
\subsection{Dataset Repository}
We publicly disclose the dataset used in this paper at the following link: \url{https://pan.baidu.com/s/1mXzVD8tjeD0wyOS879xGWA?pwd=3rbm}. Note that we only disclose the \textbf{Amazon} dataset, since \textbf{Commercial} dataset is currently not available for publicity because of policy and law restriction.

Also, we give the dataset example as follows.

\begin{figure}[!t]
\begin{subfigure}[h]{0.5\textwidth}
    \centering
    \includegraphics[width=\textwidth]{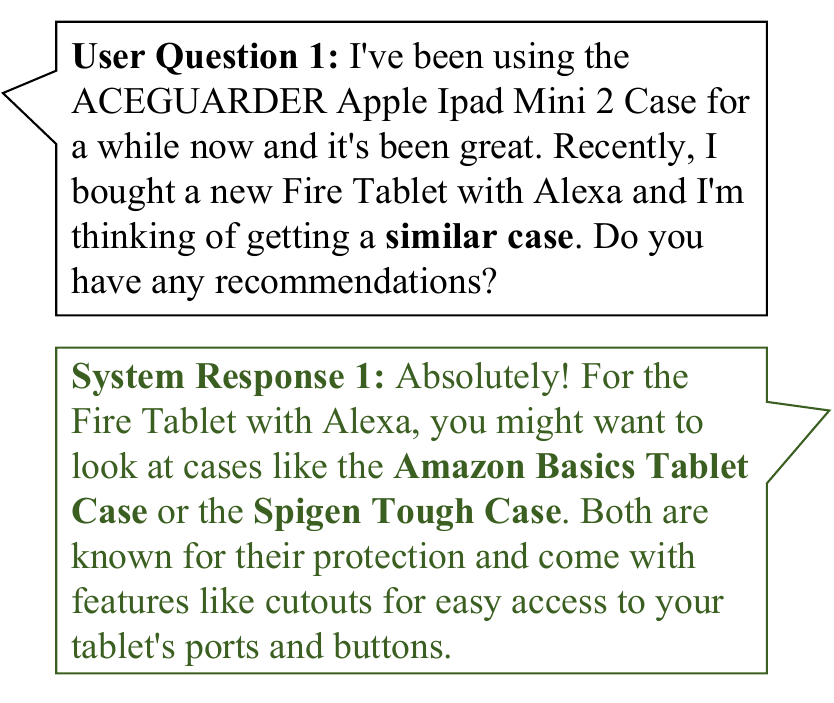}
\end{subfigure}
\begin{subfigure}[h]{0.5\textwidth}
    \centering
    \includegraphics[width=\textwidth]{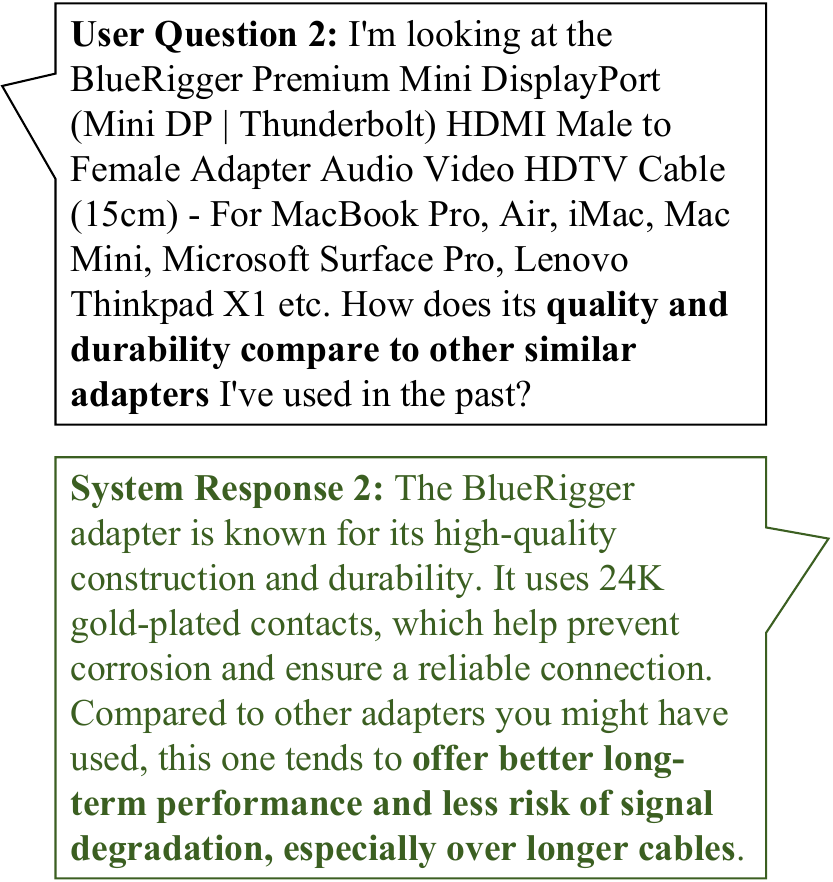}
\end{subfigure}
  \caption{Examples of consultations on the \textbf{Amazon} dataset.}
  \label{fig:app:amazon:1}
\end{figure}

\begin{figure}[!t]
\begin{subfigure}[h]{0.5\textwidth}
    \centering
    \includegraphics[width=\textwidth]{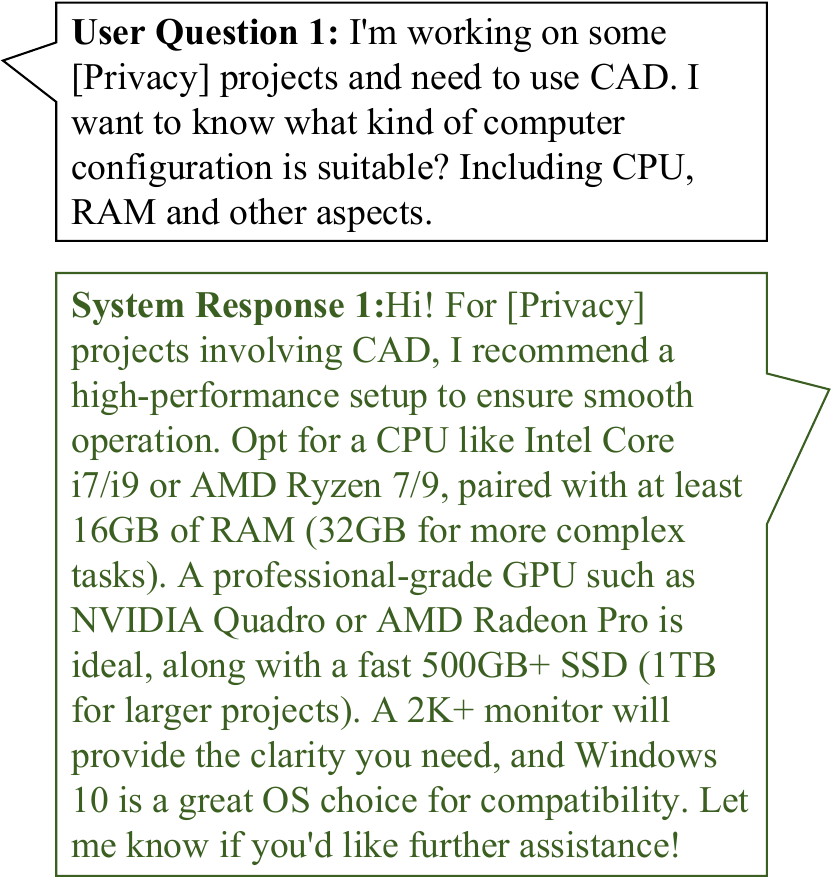}
\end{subfigure}
\begin{subfigure}[h]{0.5\textwidth}
    \centering
    \includegraphics[width=\textwidth]{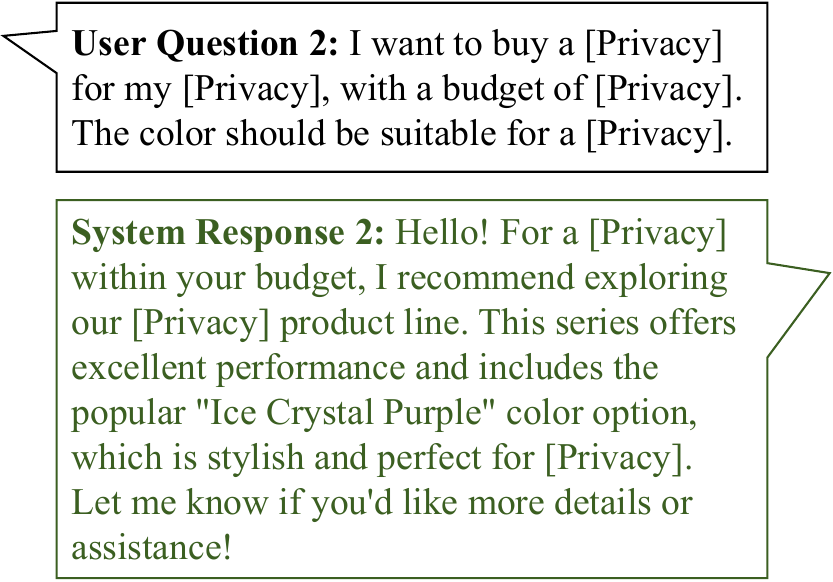}
\end{subfigure}
  \caption{Examples of consultations on the \textbf{Commercial} dataset.}
  \label{fig:app:commercial:1}
\end{figure}

\end{document}